\documentclass{astlb}

\usepackage[hyperfootnotes=false]{hyperref}
\usepackage{graphicx}
\usepackage{natbib}
\usepackage{color}

\usepackage{multirow}

\definecolor{darkblue}{rgb}{0,0,0.9}

\newcommand{\beq}{\begin{equation}}
\newcommand{\eeq}{\end{equation}}
\newcommand{\beqa}{\begin{eqnarray}}
\newcommand{\eeqa}{\end{eqnarray}}

\newcommand{\Ts}{T_{\rm s}}
\newcommand{\Tigm}{T_{\rm IGM}}
\newcommand{\Tcmb}{T_{\rm CMB}}

\newcommand{\Csfr}{C_{\rm sfr}}
\newcommand{\Lx}{L_{\rm X}}
\newcommand{\exnow}{\epsilon_{\rm X,0}}
\newcommand{\Geff}{\Gamma_{\rm eff}}
\newcommand{\ex}{\epsilon_{\rm X}}
\newcommand{\Cx}{C_{\rm X}}

\newcommand{\NH}{N_{\rm{H}}}

\newcommand{\xion}{x_{\rm i}}

\newcommand{\Rmean}{\bar{R}}
\newcommand{\Lmean}{\bar{\lambda}}

\newcommand{\rvir}{r_{\rm vir}}
\newcommand{\fgas}{f_{\rm gas}}
\newcommand{\mpr}{m_{\rm p}}
\newcommand{\Emin}{E_{\rm min}}
\newcommand{\Emax}{E_{\rm max}}

\newcommand{\nh}{n_{\rm H}}

\newcommand{\xhi}{x_{\rm HI}}
\newcommand{\nhi}{n_{\rm HI}}
\newcommand{\sigmahi}{\sigma_{\rm HI}}
\newcommand{\Gammahi}{\Gamma_{\rm HI}}
\newcommand{\Ihi}{I_{\rm HI}}

\newcommand{\xhii}{x_{\rm HII}}
\newcommand{\nhii}{n_{\rm HII}}
\newcommand{\alphahii}{\alpha_{\rm HII}}

\newcommand{\nhe}{n_{\rm He}}

\newcommand{\xhei}{x_{\rm HeI}}
\newcommand{\nhei}{n_{\rm HeI}}
\newcommand{\sigmahei}{\sigma_{\rm HeI}}
\newcommand{\Gammahei}{\Gamma_{\rm HeI}}
\newcommand{\Ihei}{I_{\rm HeI}}

\newcommand{\xheii}{x_{\rm HeII}}
\newcommand{\nheii}{n_{\rm HeII}}
\newcommand{\sigmaheii}{\sigma_{\rm HeII}}
\newcommand{\Gammaheii}{\Gamma_{\rm HeII}}
\newcommand{\alphaheii}{\alpha_{\rm HeII}}
\newcommand{\Iheii}{I_{\rm HeII}}

\newcommand{\xheiii}{x_{\rm HeIII}}
\newcommand{\nheiii}{n_{\rm HeIII}}
\newcommand{\alphaheiii}{\alpha_{\rm HeIII}}

\newcommand{\nel}{n_{\rm e}}

\newcommand{\nshi}{N_{\rm s,HI}}
\newcommand{\nshei}{N_{\rm s,HeI}}
\newcommand{\fheat}{f_{\rm heat}}

\newcommand{\rhob}{\rho_{\rm b}}

\newcommand{\Hnow}{H_{\rm 0}}
\newcommand{\Omegam}{\Omega_{\rm m}}
\newcommand{\Omegal}{\Omega_\Lambda}
\newcommand{\Omegab}{\Omega_{\rm b}}

\newcommand{\sigmaT}{\sigma_{\rm T}}
\newcommand{\sigmaSB}{\sigma_{\rm SB}}
\newcommand{\mel}{m_{\rm e}}
\newcommand{\thub}{t_{H}}

\begin{document}

\journalinfo{2017}{43}{4}{1}[0]

\title{Preheating of the early Universe by radiation from high-mass X-ray binaries}

\author{S.~Sazonov \email{sazonov@iki.rssi.ru}\address{1},
  I.~Khabibullin\address{1,2}, \addresstext{1}{Space Research
    Institute, Moscow, Russia} \addresstext{2}{Max Planck Institute
    for Astrophysics, Garching, Germany}
}
\shortauthor{Sazonov \& Khabibullin}
\shorttitle{Preheating of the early Universe by HMXBs}
\submitted{October 14, 2016}

\begin{abstract}
Using a reliably measured intrinsic (i.e. corrected for absorption effects) present-day luminosity function of high-mass X-ray binaries (HMXBs) in the 0.25--2~keV energy band per unit star-formation rate, we estimate the preheating of the early Universe by soft X-rays from such systems. We find that X-ray irradiation, mainly executed by ultraluminous and supersoft ultraluminous X-ray sources with luminosity $\Lx\ga 10^{39}$~erg~s$^{-1}$, could significantly heat ($T>\Tcmb$, where $\Tcmb$ is the temperature of the cosmic microwave background) the intergalactic medium by $z\sim 10$ if the specific X-ray emissivity of the young stellar population in the early Universe was an order of magnitude higher than at the present epoch (which is possible due to the low metallicity of the first galaxies) and the soft X-ray emission from HMXBs did not suffer strong absorption within their galaxies. This makes it possible to observe the 21~cm line of neutral hydrogen in emission from redshifts $z\la 10$. 

\englishkeywords{early Universe, reionization, high-mass X-ray binaries, ultraluminous X-ray sources} 

\end{abstract}

%%%%%%%%%%%%%%%%%%
\section{Introduction}
\label{s:intro}
%%%%%%%%%%%%%%%%%%

It is now well established (see, e.g., \citealt{fanetal06}) that nearly all hydrogen in the intergalactic space of the Universe became reionized a billion years (at redshift $z\sim 6$) after its recombination at $z\sim 1100$. Most likely, the gas was photoionized by ultraviolet radiation from the first galaxies and quasars. However, the history of galaxy formation, supermassive black hole growth and ionization of the intergalactic medium (IGM) in the early Universe (at $z>6$) remains poorly known.

Search for distant galaxies and quasars plays the most important role in the study of the early Universe. The Hubble Space Telescope has discovered more than a thousand galaxy candidates at $z\sim 6$--10 \citep{bouetal15}, which made it possible to approximately describe the history of star formation in the Universe from $\sim 500$ million to $\sim 1$ billion years after the Big Bang (see, \citealt{stark16} for review). However, current surveys are capable of finding only rare massive galaxies at such huge cosmological distances, although according to the hierarchical paradigm of structure formation there must have been much larger numbers of dwarf galaxies in the early Universe and it is the stellar population of such objects that, most likely, was mainly responsible for the reionization of the Universe.  

Virtually the only way to directly probe the IGM during the reionization epoch at $z>6$ is observation of the 21~cm line of neutral hydrogen, shifted into the meter waveband owing to the expansion of the Universe. One of the main goals of the largest radiointerferometer of the next generation -- the Square Kilometer Array,
SKA\footnote{https://www.skatelescope.org} -- is to search for a 21~cm signal from the reionization epoch. Future observations may detect neutral gas (HI) between HII regions arising around galaxies in the early Universe. It is believed that as new stars (and black holes) kept forming in the Universe, the volume fraction of HI gradually declined from $>99.9$\% at $z\sim 30$ (when the first stars appeared) to $<1$\% at $z\sim 6$. Therefore, taking into account that
the observed wavelength of the spin-flip transition of neutral hydrogen depends on redshift as $\lambda=21 (1+z)$~cm, mapping the sky in the meter waveband  enables one to trace the reionization of the Universe.

The intensity of emission or absorption in the 21~cm line is determined not only by the volume fraction of neutral hydrogen, but also by its spin temperature, $\Ts$, which in turn is established as a result of interaction of HI atoms with surrounding atoms, electrons, and photons (see \citealt{priloe12} for review). It is expected that upon the appearance of a significant UV (at energies above the Ly$\alpha$ transition and below the hydrogen ionization threshold, i.e. between 10.2 and 13.6~eV) radiation field from the first stars at $z\sim 30$--20, the HI spin temperature became (as a result of resonant scatterings, \citealt{wouthuysen52,field58}) nearly equal to the kinetic temperature of the gas, $\Tigm$. Therefore, by measuring the intensity of the cosmological 21~cm signal and its dependence on redshift at $z\la 30$, one can obtain information on the temperature of the intergalactic quasi-neutral (outside HII zones) medium at the initial stages of cosmic reionization. 

Over the last 15 years, the hypothesis has been actively discussed that the intergalactic gas could be substantially heated (above the temperature of the cosmic microwave background, CMB, $\Tcmb=2.726 (1+z)$~K) already by $z\sim 15$--10 as a result of its partial ionization by radiation from the first X-ray sources 
\citep{venetal01,madetal04,ricost04,miretal11,fraetal13b,powetal13,kneetal14,xuetal14,fiaetal14,madfra16} and low-energy cosmic rays from the first supernovae
\citep{sazsun15}. In contrast to the ionizing UV radiation (at energies above 13.6~eV), which is confined within HII zones, X-rays are capable of freely propagating and ionizing hydrogen and helium over the entire Universe. The most natural candidate for the role of X-ray sources in the early Universe is high-mass X-ray
binaries (HMXBs), since such objects provide the main contribution to the X-ray emission of actively star-forming galaxies (but without an active nucleus), according to existing observations at $0<z < 5$ \citep{lehetal16}.

It is thus possible that the gas temperature, $\Tigm(z)$, in the early Universe and, consequently, the expected intensity of emission (if $\Tigm>\Tcmb$) or absorption (if $\Tigm<\Tcmb$) in the 21~cm line were mainly determined by the integrated luminosity of the  HMXBs present at that epoch. Important is only the luminosity in the soft X-ray band (below 2~keV), because harder radiation is not capable of heating the ambient medium efficiently, owing to the rapidly declining with energy cross-section of photoabsorption on hydrogen and helium. Therefore, apart from the abundance of X-ray binaries in the early Universe, another key factor in X-ray heating is the spectrum of the emission from such objects.

In our recent study \citep{sazkha17}, we measured the specific (per
unit star-formation rate, SFR) luminosity function of HMXBs in the
local Universe in the 0.25--2~keV energy band. Besides the softer
energy range, this luminosity function is also different from those
published before (in particular, \citealt{minetal12}) in that it is
corrected for effects of absorption in the interstellar medium (ISM)
of HMXB host galaxies. We also estimated the relative contributions of
sources of various X-ray spectral types to the luminosity function. It
turned out that: i) the main contribution to the collective emission
of the HMXB population is provided by the most powerful sources with
luminosity higher than $\sim 10^{39}$~erg/s, i.e. ultraluminous X-ray
sources (ULXs), ii) nearly two thirds of the total energy release in
the soft X-ray band is provided by sources with soft and supersoft
spectra, i.e. those for which 60 and 95\%, respectively, of the total X-ray (0.25--8~keV) luminosity is emitted in the 0.25--2~keV range. The goal of the present study is to estimate the preheating of the early Universe by HMXB radiation based on the statistical properties of such objects at the present epoch explored in \cite{sazkha17}.

We use the following values of cosmological parameters in our calculations: $\Omegam=0.309$, $\Omegal=1-\Omegam$, $\Omegab=0.049$, $\Hnow=68$~km~s$^{-1}$~Mpc$^{-1}$, $\Tcmb (z=0)=2.726$~K, and $Y=0.246$ (helium mass fraction).

%%%%%%%%%%%%%%%%%%%%%%%%%
\section{Heating treatment}
\label{s:calc}
%%%%%%%%%%%%%%%%%%%%%%%%%

To calculate the heating of the early Universe by HMXB radiation it is necessary to know the following: i) redshift dependence of the SFR in the Universe
(per unit volume), ii) specific (per unit SFR) luminosity of the HMXB population in the soft X-ray band (which may depend on the metallicity in star-formation regions, see below), and iii) the fraction of the radiative energy output of X-ray sources that goes into heating of the ambient gas. 

%%%%%%%%%%%%%%%%%%%%%%%%%%%%%%%%%%%%%%%%%%%%%%%%%%%
\subsection{Cosmic star formation history}
\label{s:sfr}
 
Thanks to ultra-deep extragalactic surveys conducted recently with the Hubble Space Telescope's Wide Field Camera~3 \citep{bouetal15}, it has for the first time become possible to study the history of star formation in the early Universe ($z>6$) using direct observations of that epoch (see \citealt{maddic14,stark16} for
review). \cite{madfra16} have presented an updated dependence of the total SFR per unit volume (in comoving coordinates) in the Universe as a function of redshift:
\beq
\psi(z)=0.01~\Csfr\frac{(1+z)^{2.6}}{1+[(1+z)/3.2]^{6.2}}~M_\odot~{\rm
  yr}^{-1}~{\rm Mpc}^{-3}. 
\label{eq:sfr}
\eeq
This expression represents well the existing observational data in the $4\la z\la 10$ range. Unfortunately, there are no such data for yet earlier epochs ($z>10$) . However, since the rapid decline of the SFR is expected to continue with increasing redshift, it seems reasonable to use equation~(\ref{eq:sfr}) also at $z>10$ (so that $\psi(z)\propto (1+z)^{-3.6}$) in our calculations.

Equation~(\ref{eq:sfr}) was obtained using observations of distant galaxies with fairly high luminosity, by extrapolating the measured galaxy luminosity function into the currently poorly studied region of lower luminosities (i.e. more numerous small galaxies). This extrapolation, as well as the aforementioned lack of observations of galaxies at $z>10$, causes a substantial uncertainty, which can be crudely described by a coefficient, $\Csfr$, of the order of unity in equation~(\ref{eq:sfr}). 

%%%%%%%%%%%%%%%%%%%%%%%%%%%%%%%%%%%%%%%%%%%%%%%%%%%%%%%%%%%%%%%%
\subsection{Specific X-ray luminosity of the young stellar population in the early Universe} 
\label{s:lx_sfr}

In \citep{sazkha17} we have obtained an approximate analytic expression for the intrinsic (i.e. corrected for absorption in the ISM of the Galaxy and host galaxies, as well as for related source selection effects) luminosity function of HMXBs in nearby galaxies in the 0.25--2~keV energy band and luminosity range of $\Lx=10^{38}$--$10^{40.5}$~erg~s$^{-1}$, per unit SFR:
\beq
\frac{dN}{d\log\Lx}=A(\Lx/10^{39}~\mathrm{erg~s}^{-1})^{-\gamma},
\label{eq:xlf}
\eeq
where $A=(1.36\pm 0.15)$~($M_\odot$~yr$^{-1}$)$^{-1}$ and  $\gamma=0.63\pm0.08$.   

We also demonstrated that nearly equal contributions to the luminosity function are provided by hard, soft and supersoft sources, defined as those with a 0.25--2~keV to 0.25--8~keV luminosity ratio of $<0.6$, 0.6--0.95, and $>0.95$, respectively. Therefore, more than half of the total unabsorbed soft X-ray emission of star forming galaxies at the present epoch is produced by HMXBs with soft X-ray spectra. Furthermore, as follows from the shallow slope of the luminosity function, the main contribution is provided by sources with $\Lx\ga 10^{39}$~erg~s$^{-1}$, i.e. ultraluminous X-ray sources (ULXs) and (in the case of a very soft spectrum) ultraluminous supersoft X-ray sources. Most likely (see \citealt{urqsor16} and a discussion in \citealt{sazkha17}), the majority of such sources are HMXBs with supercritical accretion onto a stellar-mass black hole, with hardness or softness of the measured spectrum being determined by the orientation of the supercritical disk, with an outflowing wind, with respect to the observer.  

By integrating the luminosity function given by equation~(\ref{eq:xlf}) over the $\Lx$ range from $10^{38}$ to $10^{40.5}$~erg~s$^{-1}$, we obtain the cumulative luminosity of the young stellar population at $z=0$ per unit SFR in the 0.25--2~keV energy band:
\beqa
\exnow&\equiv& \int\frac{dN}{d\log\Lx}\Lx d\log\Lx \\
&\approx& 5.0\times 10^{39}~{\rm erg~s}^{-1}~(M_\odot~{\rm yr}^{-1})^{-1}. 
\label{eq:exnow}
\eeqa

As was also shown in \citep{sazkha17}, the cumulative luminosity in the 0.25--2~keV range is approximately two times the corresponding luminosity in the harder range of 2--8~keV. Therefore, the effective slope (photon index) of the spectrum of the collective X-ray emission of HMXBs is:
\beq
\Geff\approx 2.1.
\label{eq:gamma_eff}
\eeq

A number of recent studies \citep{basetal13,broetal14,douetal15,basetal16,lehetal16} indicate that in galaxies with low metallicity the specific X-ray luminosity $\ex$ is an order of magnitude higher than in galaxies with normal (solar) chemical composition. Although these observational results are characterized by significant uncertainty and need further verification, there are substantial reasons to expect the occurence rate and X-ray luminosities of HMXBs to increase with decreasing metallicity: lower abundance of heavy elements can lead to weakened matter outflow from OB and Wolf--Rayet stars, and consequently enhanced formation of black holes (relative to neutron stars) and their increased average mass, as well as to formation of tighter binary systems in which intensive  accretion onto the black hole can take place as a result of overflow of the Roche lobe by the massive stellar companion. This problem has been actively discussed recently and population synthesis modeling indeed indicates (see, e.g., \citealt{linetal10,fraetal13a}) that lowering the metallicity to $Z\la 0.1Z_\odot$ can be accompanied by an order of magnitude increase in $\ex$ relative to the $Z=Z_\odot$ case.

Because in the epoch of interest here ($z<6$) the metallicity was probably low ($Z\la 0.1Z_\odot$) even in star-forming regions (see, e.g., \citealt{paletal14}), we may assume that the specific X-ray (0.25--2~keV) luminosity of the young stellar population exceeded the value given by equation~(\ref{eq:exnow}):
\beq
\ex=\Cx\exnow=5\times 10^{39}\Cx~{\rm erg~s}^{-1}~(M_\odot~{\rm yr}^{-1})^{-1},
\label{eq:ex}
\eeq
where $\Cx$ can range between 1 and $\sim 10$. 

%%%%%%%%%%%%%%%%%%%%%%%%%%%%%%%%%%%%%%%%%%%%%%%%%%%%%%%%%%%%%%%%%
\subsection{Efficiency of gas ionization and heating by X-rays} 
\label{s:heateff}

As was mentioned in many previous works, X-ray heating of quasi-neutral gas with primordial chemical composition mainly proceeds via photoionization of helium atoms despite the fact that their number is just $\sim 8$\% of that of hydrogen atoms. This is due to the fact that the cross-section for photoionization of X-ray photons by HeII is nearly 30 times that for HI \citep{veretal96}. As a result of ionization of an atom of helium or hydrogen by an X-ray photon, a fast free electron is produced, which shortly shares its kinetic energy with the surrounding gas. However, this energy is expended not only on gas heating (as a result of collisions of the fast electron with thermal ones) but also on ionization and excitation of other helium and hydrogen atoms. The mean fraction of X-ray photon energy that goes into heating depends weakly on the photon energy $E$ if $E\ga 250$~eV (as in the situation under consideration), but strongly on the gas ionization degree (fraction of free electrons) $\xion$, rising from $\sim 12$\% at $\xion\sim 10^{-4}$ to  $\sim 17$\% at $\xion=10^{-3}$, $\sim 35$\% at $\xion=10^{-2}$, and $\sim
70$\% at $\xion=0.1$ \citep{fursto10}. 

At the present epoch, the soft X-ray radiation (0.25--2~keV) from HMXBs experiences strong absorption in the ISM (atomic and molecular gas) of their host galaxies. In the first galaxies, absorption could  be much weaker due to the low metallicity of their ISM. Indeed, a typical absorption column density in the direction of HMXBs in nearby galaxies is $\NH\sim 3\times 10^{21}$~H~atoms per sq. cm \citep{sazkha17}, with $Z\sim Z_\odot$. In such a situation, only $\exp(-\sigma_{\rm ph}(E)\NH)\sim 45$\% of photons with $E\sim 1$~keV and $\sim 10$\% of photons with $E\sim 0.5$~keV (where $\sigma_{\rm ph}$ is the total photoionization cross-section per hydrogen atom) escape from the galaxy, whereas yet softer radiation is almost completely absorbed within the galaxy. If the first galaxies had the same surface density of neutral hydrogen but $Z\la 0.1Z_\odot$, the corresponding fractions would be $\ga 80$\% and $\ga 20$\%. In the case of somewhat lower column density $\NH\sim 10^{21}$~cm$^{-2}$ and $Z\la 0.1Z_\odot$, these fractions would be even higher: $\ga 90$\% and $\ga 60$\%. These estimates have been obtained using the \texttt{vphabs} model in {\sc xspec} and photoabsorption cross-sections from \citep{veretal96}.

In reality, however, the first galaxies were much more compact but characterized by higher average gas density. It is difficult to reliably estimate the typical column density in such a situation. A crude estimate can be obtained by assuming that the gas is distributed uniformly within the galaxy virial radius and is not ionized by UV radiation from stars and supernovae. A halo of mass $M$ collapsing at redshift $z$ has a virial radius $\rvir\sim 1.5 (M/10^8M_\odot)^{1/3}[(1+z)/10]^{-1}$~kpc. Assuming that the gas mass fraction within the halo is equal to the average baryonic fraction in the Universe ($\fgas=\Omegab/\Omegam\sim 0.16$), we find that $\NH\sim \fgas (1-Y) M/(4/3\pi\rvir^2\mpr)\sim 1.5\times 10^{20}(M/10^8M_\odot)^{1/3}[(1+z)/10]^{2}$~cm$^{-2}$. Since typical galaxies at $z\sim 10$ have total masses $\ga 10^8M_\odot$, we conclude that the (low-metallicity) ISM in such objects was characterized by surface density comparable to or somewhat
lower than in present-day galaxies. 

Therefore, a sizeable but likely lower than at $z=0$ fraction of the soft X-ray radiation from HMXBs might have been able to escape from the galaxies in the early Universe. Taking into account the significant uncertainty associated with the magnitude of this absorption, we can assume that the spectrum of X-ray radiation emergent from the first galaxies has a sharp boundary at energy $\Emin\sim 0.5$~keV, which may be regarded as a parameter in the following calculations.  

%%%%%%%%%%%%%%%%%%%%%%%%%%%%%%%%%%%%%%%%%%%%%%%%%%%%%%%%%%%%%%%%%%%%%%%%
\subsection{Spatial and timing properties of X-ray heating}
\label{s:heatprop}

As was discussed above, the main contribution to the collective X-ray emission of the HMXB population is provided by ULXs with $\Lx\ga 10^{39}$~erg~s$^{-1}$. In the local Universe, the occurence rate of such objects is $\sim 1$ per a SFR of $M_\odot$~yr$^{-1}$ (see equation  
(\ref{eq:xlf})). Consequently, the average (proper) distances between such objects in the early Universe are expected to be $\Rmean\sim 0.7\Cx^{-1/3}$~Mpc at $z=15$ and  $\sim 0.6\Cx^{-1/3}$~Mpc at $z=6$, where we have used the dependence of the SFR on redshift (\ref{eq:sfr}) and taken into account that the abundance of HMXBs in the early Universe could be $\Cx$ times higher than at $z=0$.

On the other hand, the mean free path of X-ray photons in the early Universe can be approximately described by the relation \citep{sazsun15}
\beq
\Lmean\sim 5\left(\frac{1+z}{10}\right)^{-3}\left(\frac{E}{500~{\rm
    eV}}\right)^{3.2}\,{\rm Mpc},
\label{eq:path_appr}
\eeq
where we have taken into account the dependence of the HI and HeI photoionization cross-sections on energy and the dependence of the baryonic density of the Universe on redshift. Therefore, $\Lmean> \Rmean$ for  $E\ga 300$~eV. We can thus expect IGM heating in the early Universe to mainly take place as a result of irradiation by soft X-rays from numerous ULXs. At the same time, it is obvious that heating was particularly strong near such sources, which will be interesting to study in future work. 

\begin{figure}
\centering
\includegraphics[width=\columnwidth,viewport=20 180 560 720]{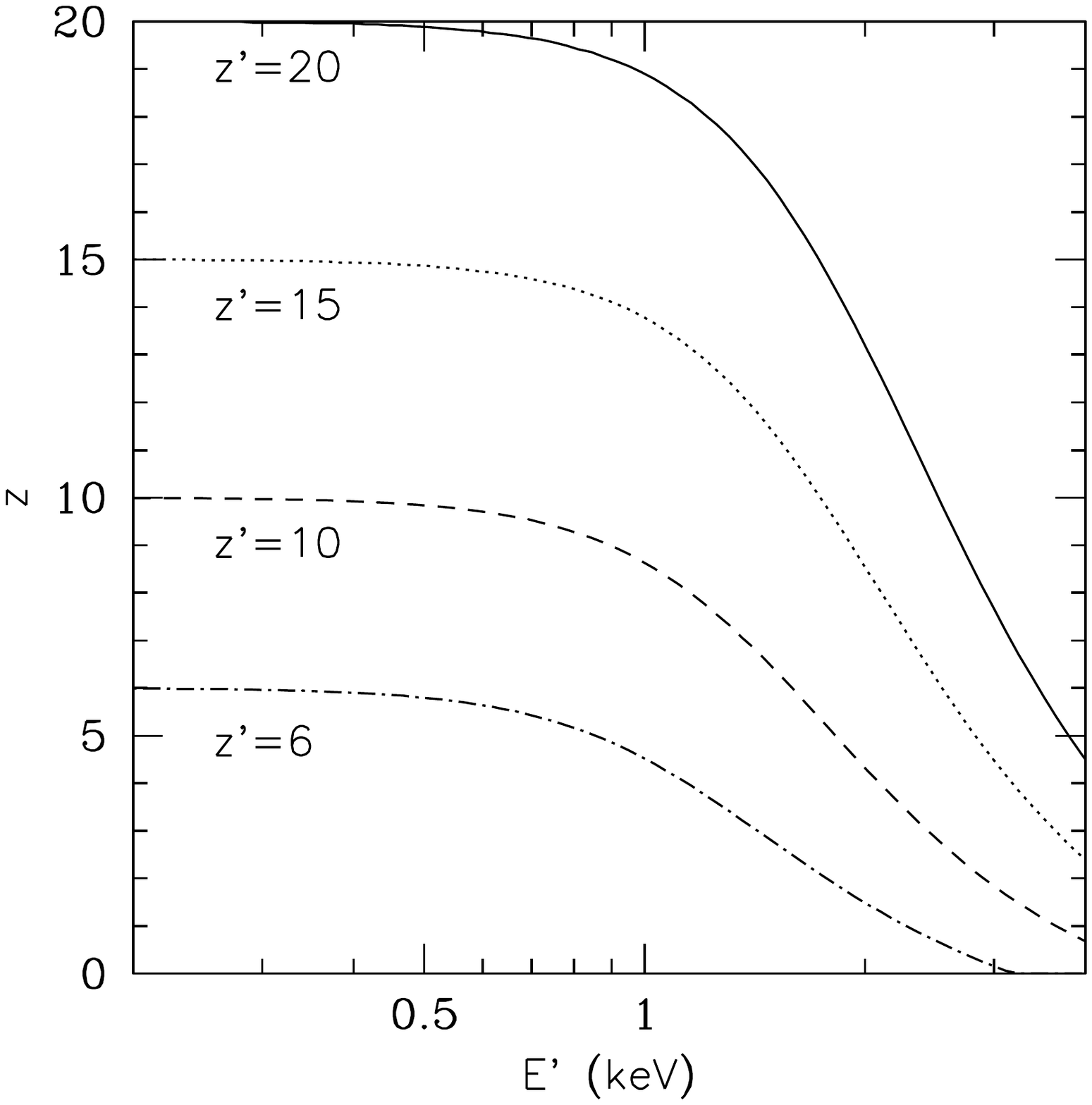}
\caption{Redshift, $z$, at which a photon emitted at redshift $z'$ with initial energy $E'$ is expected to be photoabsorbed.}
\label{fig:e_zin_zfi}
\end{figure}

According to equation~(\ref{eq:path_appr}), $\Lmean/c\ll 1/H (z)$ -- the characteristic time of the expansion of the Universe at a given $z$ -- in the redshift range of interest  $z\sim 15$--6 for photon energies $E\la 1$~keV. In considering the fate of higher energy photons, we need to take the expansion of the Universe into account, which causes the photon energy and IGM density to decrease with time. Figure~\ref{fig:e_zin_zfi} shows the result of an accurate calculation of the redshift, $z$, at which the photoabsorption of a photon emitted at redshift $z'$ with initial energy $E'$ is expected (the optical depth of the Universe between $z'$ and $z$ for such a photon is equal to unity). The presented curves have been obtained for a gas ionization degree of $\xion=2\times 10^{-4}$, but they remain practically unchanged for $\xion\la 0.1$. Note that in our calculation the chemical composition of the IGM was assumed to be primordial (hydrogen and helium), although at $z\la 6$ heavy elements could have already spread in significant amounts over the Universe, which must have accelerated absorption of X-ray photons. Therefore, the $z\la 6$ parts of the curves in Fig.~\ref{fig:e_zin_zfi} are likely to be inaccurate. This is however not important for this study, since we are interested in events taking place in the Universe at $z>6$.  

As follows from this figure, IGM heating by photons with $E\la 500$~eV can be considered instantaneous, while heating by harder X-rays proves to be substantially spread over time, with $E\ga 2$~keV photons being practically incapable of exchanging their energy with the gas (taking into account that such photons manage to lose most of their energy adiabatically before being absorbed in the IGM). 

%%%%%%%%%%%%%%%%%%%%%%%%%%%%%%%%
\subsection{Calculation method}
\label{s:algorithm}

We carried out a calculation of IGM ionization and heating in the early Universe using (with slight modifications) the formalism described in the recent work \citep{madfra16}. Specifically, the average intensity of the cosmic X-ray background produced by HMXBs at $z$ is
\beq
J_{E}=\frac{c}{4\pi}(1+z)^3\int_z^{z_0}\frac{dz'}{(1+z')H(z')}\epsilon(E')e^{-\tau(E',z')},
\label{eq:intens}
\eeq
where $z_0$ is the redshift of the epoch when the first HMXBs appeared, $E'=E(1+z')/(1+z)$, and $\tau(E',z')$ is the optical depth between $z'$ and 
$z$ for a photon with initial energy $E'$, which can be described as follows:
\beq
\tau(E',z')=c\int_z^{z'}\frac{d\tilde{z}}{(1+\tilde{z})H(\tilde{z})\lambda(\tilde{E},\tilde{z})},
\eeq
where $\lambda$ is the photon mean free path:
\beq
\lambda(\tilde{E},\tilde{z})=\frac{1}{\nhi\sigmahi+\nhei\sigmahei+\nheii\sigmaheii},
\label{eq:path}
\eeq
where $\nhi(\tilde{z})$, $\nhei(\tilde{z})$, and $\nheii(\tilde{z})$ are the number densities of HI, HeI, and HeII at redshift $\tilde{z}$, and $\sigmahi(E)$, $\sigmahei(\tilde{E})$, and $\sigmaheii(\tilde{E})$ are the corresponding photoabsorption cross-sections for a photon with energy $\tilde{E}$, adopted from \citep{veretal96}. The preceeding formula (\ref{eq:path_appr}) is an approximation of equation~(\ref{eq:path}). 

The quantity $\epsilon(E)$ entering equation (\ref{eq:intens}) is the volume emissivity of the Universe at energy $E$ in comoving coordinates. According to the preceeding consideration, its integral over energy in the $E=0.25$--2~keV band is 
\beq
\int_{0.25~{\rm keV}}^{2~{\rm keV}}\epsilon(E)dE=\psi(z)\ex,
\label{eq:epsilon}
\eeq
where the SFR $\psi(z)$ and specific X-ray luminosity of the young stellar population $\ex$ are described by equations~(\ref{eq:sfr}) and (\ref{eq:ex}), respectively.  We used in our computations spectra of power-law type:
\beq
\epsilon(E)=AE^{-\Gamma+1},
\label{eq:spec}
\eeq
where $\Gamma$ was considered a model parameter and the normalizing constant $A$ was determined from the condition~(\ref{eq:epsilon}). In addition, we varied  $\Emin$ and $\Emax$, the boundaries of the energy band for the X-ray radiation escaping from the galaxies.  

The evolution of the ionization state of hydrogen ($\xhi=\nhi/\nh$, $\xhii=\nhii/\nh$) and helium ($\xhei=\nhei/\nhe$, $\xheii=\nheii/\nhe$, $\xheiii=\nheiii/\nhe$) with time can be described by the following equations:
\beqa
\frac{d\xhi}{dt} &=& -\xhi\Gammahi+\nel (1-\xhi)\alphahii,\nonumber\\
\frac{d\xhei}{dt} &=& -\xhei\Gammahei+\nel\xheii\alphaheii,\nonumber\\
\frac{d\xheii}{dt} &=&
-\xheii\Gammaheii+\nel\xheiii\alphaheiii-\frac{d\xhei}{dt}.\nonumber\\
&&
\label{eq:dxh}
\eeqa
Here $\nel$ is the number density of free electrons, $\alphahii$, $\alphaheii$, and $\alphaheiii$ are the coefficients of recombination (which plays a negligible role in the problem under consideration), adopted from \citep{theetal98}, and $\Gammahi$, $\Gammahei$, and $\Gammaheii$ are the ionization coefficients, which can be derived using the formulae
\beqa
\Gammahi=\int_{\Ihi}^\infty\frac{4\pi
  J_E}{E}\sigmahi[1+N_{{\rm s,HI},E-\Ihi}]dE\nonumber\\ 
  +\int_{\Ihei}^\infty\frac{4\pi
    J_E}{E}\sigmahei\frac{\nhei}{\nhi}N_{{\rm s,HI},E-\Ihei}dE\nonumber\\
  +\int_{\Iheii}^\infty\frac{4\pi
    J_E}{E}\sigmaheii\frac{\nheii}{\nhi}N_{{\rm s,HI},E-\Iheii}dE,\nonumber\\
\Gammahei =\int_{\Ihi}^\infty\frac{4\pi
  J_E}{E}\sigmahi\frac{\nhi}{\nhei}N_{{\rm s,HeI},E-\Ihi}dE\nonumber\\
  +\int_{\Ihei}^\infty\frac{4\pi
    J_E}{E}\sigmahei[1+N_{{\rm s,HeI},E-\Ihei}]dE\nonumber\\ 
  +\int_{\Iheii}^\infty\frac{4\pi
    J_E}{E}\sigmaheii\frac{\nheii}{\nhei}N_{{\rm s,HeI},E-\Iheii}dE,\nonumber\\
\Gammaheii =\int_{\Iheii}^\infty\frac{4\pi J_E}{E}\sigmahi dE,
\label{eq:gammas} 
\eeqa
where $\Ihi=13.6$~eV, $\Ihei=24.6$~eV, and $\Iheii=54.4$~eV are the ionization thresholds for HI, HeI, and HeII, and $\nshi$ and $\nshei$ are the mean numbers of secondary ionizations of HI and HeI (secondary ionization of HeII is practically unimportant) induced by the fast photoelectron, whose energy can take one of the three values $E-\Ihi$, $E-\Ihei$, or $E-\Iheii$. The dependencies of $\nshi$ and $\nshei$ on energy are adopted from \citep{fursto10}. 

The evolution of the gas temperature with time is given by
\beq
\frac{d\Tigm}{dt}=-2H\Tigm+\frac{\Tigm}{\mu}\frac{d\mu}{dt}+\frac{2\mu\mpr}{3k\rhob}\mathcal{H},
\label{eq:dtk}
\eeq
where
\beqa
\mathcal{H} = \int_{\Ihi}^\infty\frac{4\pi
  J_E}{E}(E-\Ihi)\nhi\sigmahi f_{{\rm heat},E-\Ihi} dE\nonumber\\  
+\int_{\Ihei}^\infty\frac{4\pi J_E}{E}(E-\Ihei)\nhei\sigmahei
f_{{\rm heat},E-\Ihei}dE\nonumber\\  
+\int_{\Iheii}^\infty\frac{4\pi J_E}{E}(E-\Iheii)\nheii\sigmaheii
f_{{\rm heat},E-\Iheii}dE,\nonumber\\
\label{eq:heatrate}
\eeqa
where $\rhob$ is the average baryonic density of the Universe, $\mu$ is the mean molecular weight, and $\fheat$ is the fraction of the photoelectron energy ($E-\Ihi$, $E-\Ihei$, or $E-\Iheii$) that goes into gas heating \citep{fursto10}.

We have neglected in equation (\ref{eq:dtk}) gas cooling and heating as a result of scattering of the CMB by free electrons, because the corresponding characteristic time of gas temperature change,  
\beqa
t_{{\rm CMB}} &=& \frac{3\mel c^2}{32\sigmaT\sigmaSB\Tcmb^4(z)}
\nonumber\\
&\approx & 1.2\times
10^{8}\left(\frac{\nel}{n}\right)^{-1}\left(\frac{1+z}{10}\right)^{-4}\,{\rm
  yr}
\label{eq:cmbtime}
\eeqa
(where $\sigmaT$ is the Thomson scattering cross-section and $\sigmaSB$ is the Stefan--Boltzman constant), significantly exceeds the Hubble time, 
\beq
\thub\approx 5\times 10^8\left(\frac{1+z}{10}\right)^{-3/2} {\rm yr},
\label{eq:hubtime}
\eeq
at low gas ionization degrees $\nel/n\la 0.01$ (where $n$ is the total number density of particles), as is the case here.

Similarly, in the problem at hand one can neglect Compton heating of the gas by X-ray radiation as well as radiative losses arising from collisional and recombination processes. We verified this by inserting the corresponding terms (see \citealt{madfra16} and \citealt{theetal98}) into equation (\ref{eq:dtk}) and repeating the computations. 

%%%%%%%%%%%%%%%%%%%%%%%%%%%%
\section{Calculation results}
\label{s:results}
%%%%%%%%%%%%%%%%%%%%%%%%%%%%

We started the calculation of ionization and heating at redshift $z=15$ with initial parameters $\xhii=2.2\times 10^{-4}$, $\xheii=\xheiii=0$, and $\Tigm=5.4$~K. These values were found using the RECFAST program \citep{seaetal99} and correspond to the conditions in the Universe upon recombination and adiabatic cooling of the gas. Similar initial conditions were adopted by \citep{madfra16}. The integration was continued up to $z=6$, i.e. until the end of the reionization epoch. 

\begin{figure}
\centering
\includegraphics[width=\columnwidth,viewport=20 180 560 720]{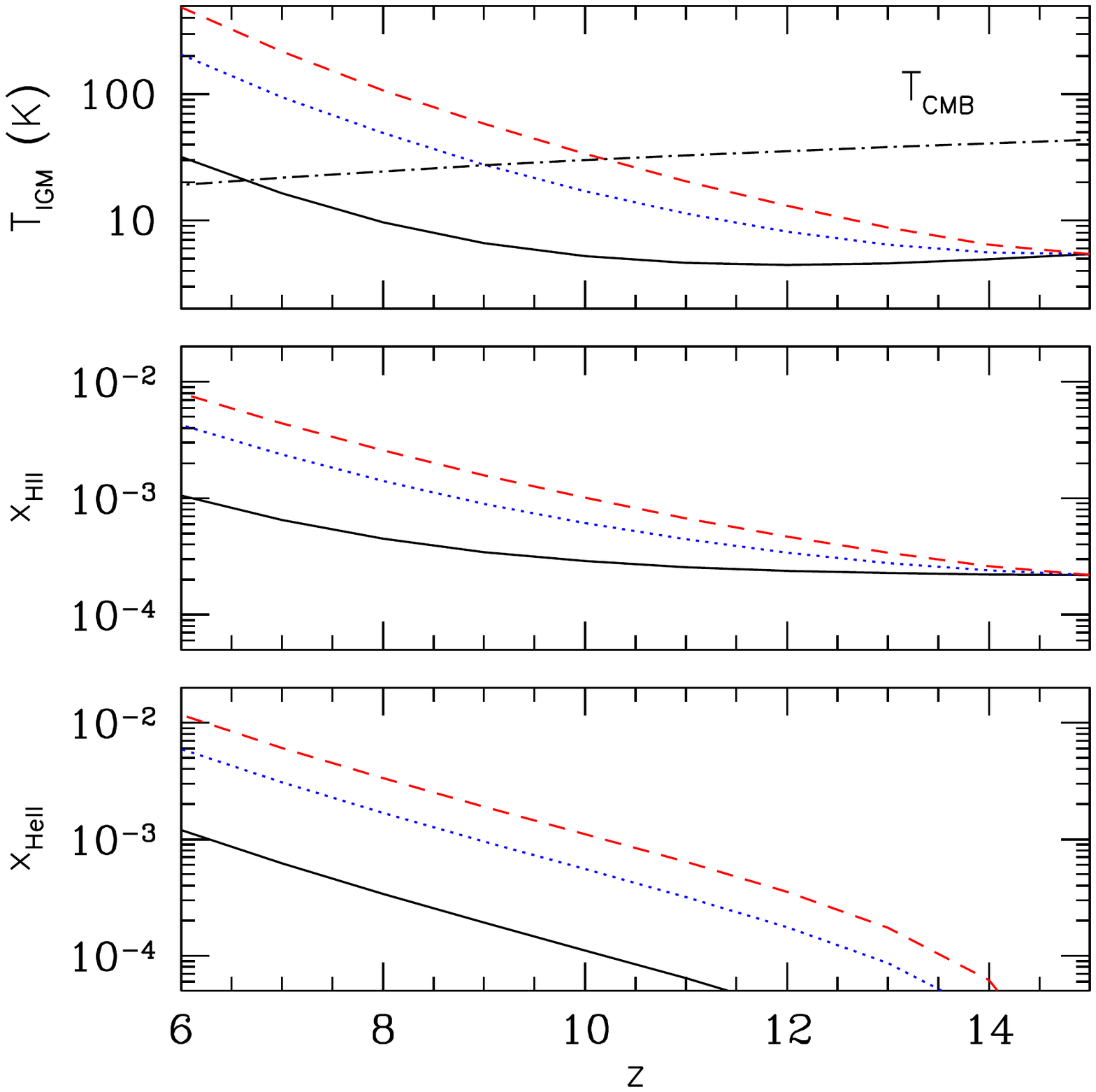}
\caption{Evolution of the IGM temperature (top panel), and the fraction of ionized hydrogen (middle panel) and singly ionized helium (bottom panel) with redshift as a result of X-ray heating by HMXBs, under the assumption that the specific X-ray luminosity of the young stellar population in the early Universe was the same as at the present epoch ($\Cx=1$, solid lines), was 5 times higher ($\Cx=5$, dotted lines), and 10 times higher ($\Cx=10$, dashed lines). It was also assumed that the HMXB emission spectrum is a power law with $\Gamma=2$ ($dL/dE\propto E^{-\Gamma+1}$) in the 0.25--10~keV energy band and that all radiation in this range can escape from the galaxies. The SFR was assumed to be standard ($\Csfr=1$). The dash-dotted line shows the evolution of the CMB temperature $\Tcmb$.
}
\label{fig:ion_heat_em}
\end{figure}

\begin{figure}
\centering
\includegraphics[width=\columnwidth,viewport=20 180 560 720]{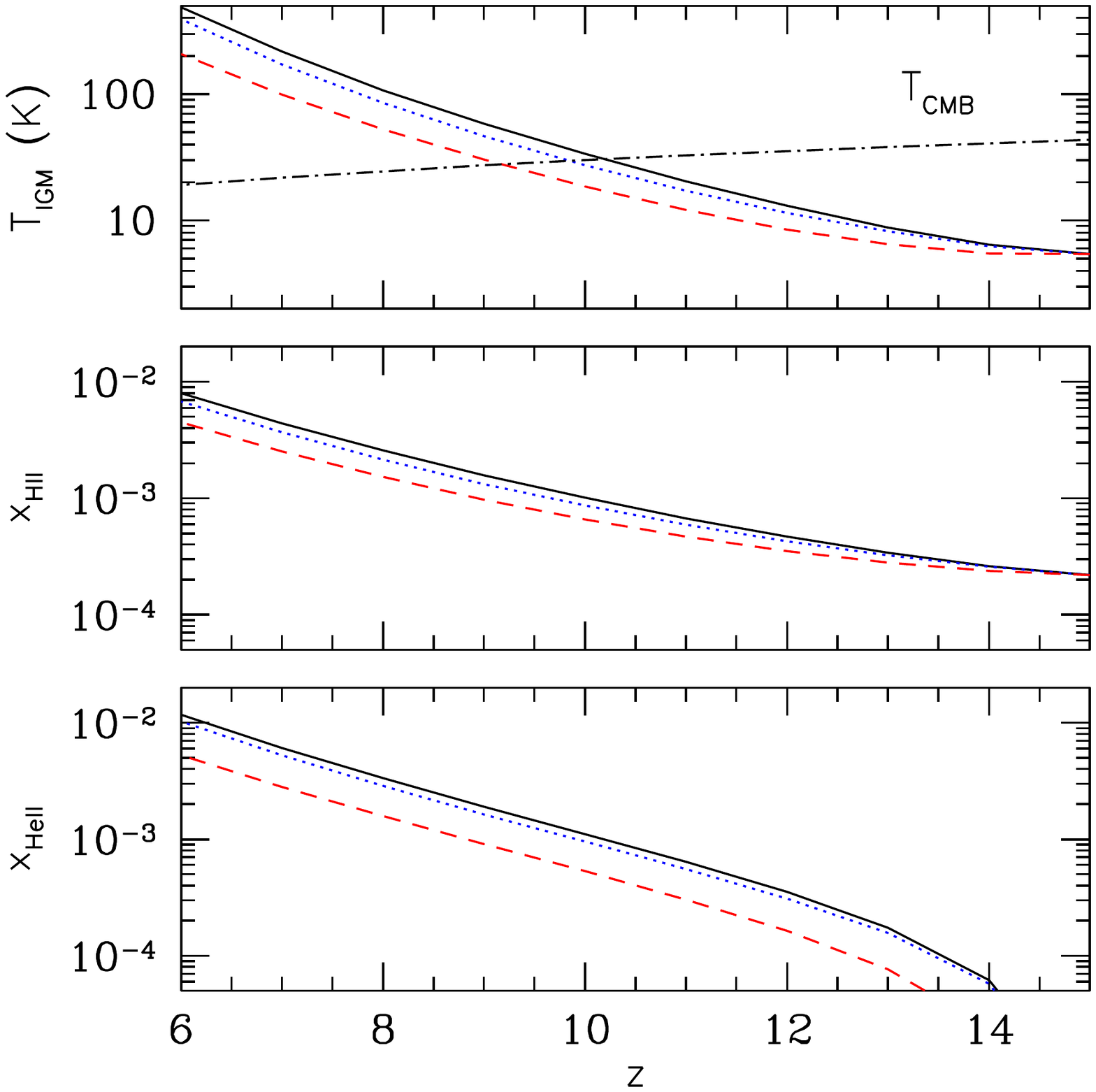}
\caption{Similar to Fig.~\ref{fig:ion_heat_em}. The parameter values are $\Cx=10$, $\Csfr=1$, and $\Gamma=2$. The different types of curves correspond to different energy bands for the X-ray radiation escaping from the galaxies: solid lines -- 0.25--10~keV, dotted lines -- 0.25--1~keV, dashed lines -- 0.5--10~keV.} 
\label{fig:ion_heat_eband}
\end{figure}

\begin{figure}
\centering
\includegraphics[width=\columnwidth,viewport=20 180 560 720]{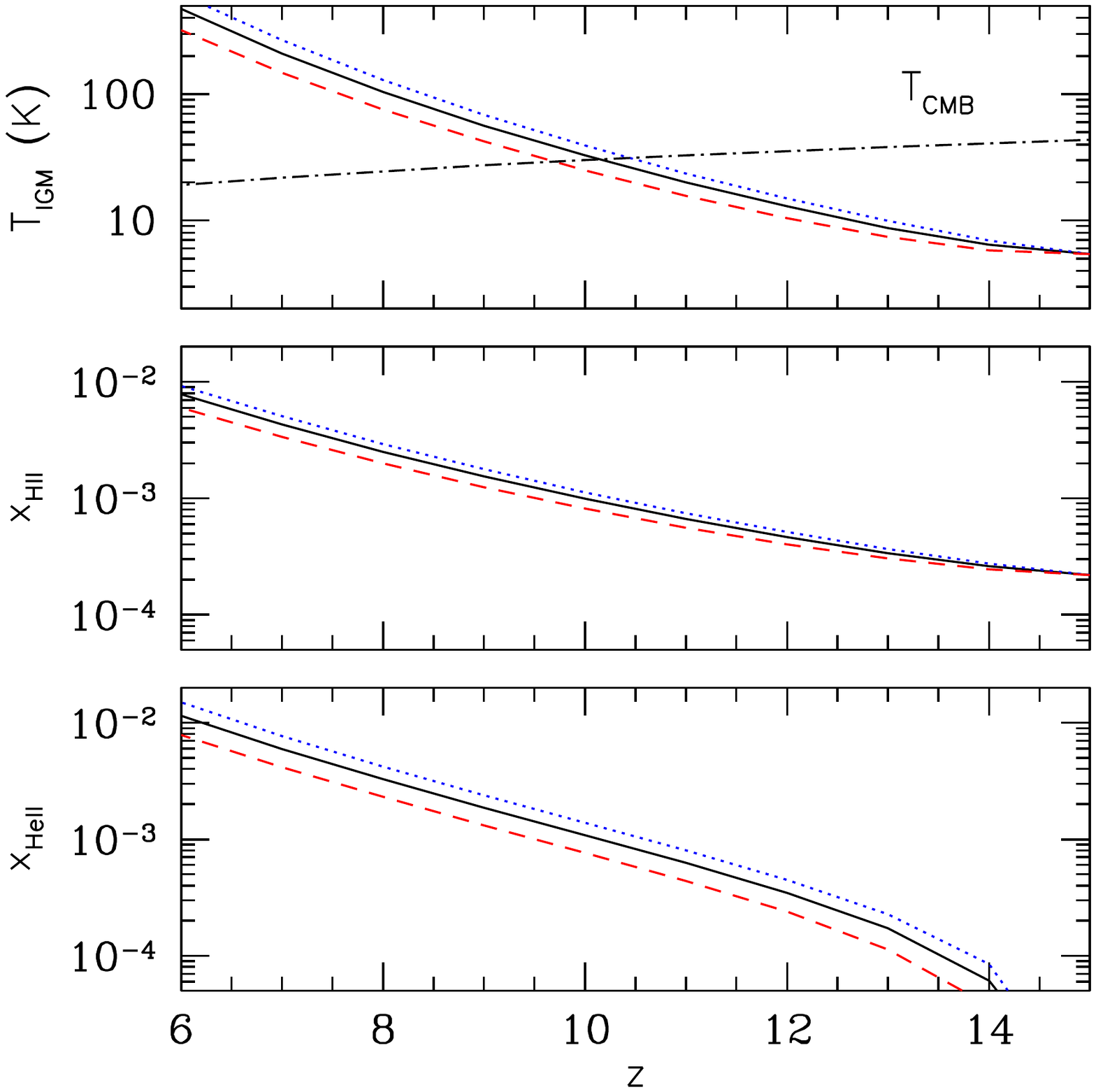}
\caption{Similar to Fig.~\ref{fig:ion_heat_eband}. The parameter values are $\Cx=10$ and $\Csfr=1$, the energy band is 0.25--2~keV. The different types of curves correspond to different spectral slopes: solid lines -- $\Gamma=2$, dotted lines -- $\Gamma=3$, dashed lines -- $\Gamma=1.$}
\label{fig:ion_heat_specslope}
\end{figure}

Figures~\ref{fig:ion_heat_em}--\ref{fig:ion_heat_specslope} show the expected dependencies $\Tigm(z)$, $\xhii(z)$, and $\xheii(z)$ (the fraction of doubly ionized helium remains negligibly small, $\xheiii<10^{-4}$, up to $z=6$ and is thus not shown) for various parameter values. It is evident from Fig.~\ref{fig:ion_heat_eband}, where the result of varying the X-ray emission energy band is shown, that X-ray photons with energy higher than 1~keV play practically no role in heating the IGM and that truncating the energy range from below at $\Emin=0.5$~keV instead of $\Emin=0.25$~keV leads to an approximately two-fold decrease of the resulting ionization and heating. Changing the slope of the input emission spectrum in the 0.25--2~keV energy band with a fixed luminosity in this range (see Fig.~\ref{fig:ion_heat_specslope}) also significantly affects the result, but to a lesser degree than varying $\Emin$. Note that the slope $\Gamma=2$ approximately corresponds to the measured ratio of HMXB integrated luminosities in the 0.25--2 and 2--8~keV energy bands \citep{sazkha17} (see equation~(\ref{eq:gamma_eff})).  

The most important role is, of course, played by the specific X-ray luminosity of the young population of the first galaxies (see Fig.~\ref{fig:ion_heat_em}). For instance, in the case of $\Cx=1$, i.e. if HMXBs in the early Universe produced as much X-ray radiation per unit SFR as at the present epoch, IGM heating proves to be weak up to $z=6$. Only in the case of substantially larger specific X-ray luminosity of HMXBs, namely $\Cx=10$ (as is in fact expected because of the low metallicity of the first galaxies), can the IGM be heated above the CMB temperature already at $z=10$, i.e. long before the Universe was ionized by the UV radiation from galaxies and quasars. However, for this scenario to be realized it is necessary that the soft X-ray radiation of HMXBs experience almost no absorption within their host galaxies (i.e. $\Emin\sim 0.25$~keV). Another possibility of strengthening the effect of X-ray heating is obviously associated with  increasing the SFR parameter $\Csfr$ above unity, which has the same effect on the result of ionization and heating as increasing $\Cx$. 

%%%%%%%%%%%%%%%%%%%%%
\section{Discussion and conclusion}
\label{s:summary}
%%%%%%%%%%%%%%%%%%%%

We have evaluated the preheating of the early Universe by soft X-ray radiation from HMXBs. In contrast to previous studies on this subject, we used a reliably measured and corrected for absorption effects specific (per unit SFR) HMXB luminosity function in the soft X-ray band of 0.25--2~keV at the present epoch 
\citep{sazkha17}.

We demonstrated that X-ray irradiation (mostly executed by ultraluminous and supersoft ultraluminous X-ray sources with luminosity $\Lx\ga 10^{39}$~erg~s$^{-1}$) could lead to a significant heating ($\Tigm>\Tcmb$) of the IGM at $z\sim 10$ (i.e. before the beginning of the active stage of reionization of the Universe by UV radiation from galaxies and quasars) only if all of the following three conditions were fulfilled: i) the SFR at $z\sim
15$--10 was such as follows from extrapolation of the observed SFR at $z\la 10$ or was higher, ii) the specific X-ray luminosity of the young stellar population of the first galaxies was an order of magnitude higher that at the present epoch, and iii) the soft X-ray radiation of HMXBs did not suffer strong absorption within their  galaxies. Only if these conditions are satisfied, can the 21~cm signal from $z\la 10$ epochs be observed in emission. The signal from the earlier epochs should be expected in absorption, since $\Tigm<\Tcmb$ at $z\ga 10$. Note, however, that the Universe could be heated to $\Tigm>\Tcmb$ at $z\ga 10$ by low-energy cosmic rays from the first supernovae \citep{sazsun15}.

It should be possible to test the first of the above conditions with the next generation of optical and infrared telescopes, which, are expected to begin  discovering galaxies at $z>10$. Fulfillment of the second condition appears plausible, as there exist observational and theoretical indications of increased abundance and X-ray luminosity of HMXBs in galaxies with low metallicity. In this connection, it is important to continue statistical studies of X-ray sources in low-metallicity galaxies in the local Universe. It is currently unclear if the third condition (about the escape of soft X-ray radiation from galaxies) is satisfied. The only way to obtain a reasonable answer to this question is to perform detailed modeling of the ISM in the first galaxies. We emphasize the importance of this aspect of the problem: raising the effective lower boundary of the spectral range from $\Emin=0.25$ to 0.5~keV leads to the attenuation of IGM heating by approximately a half, whereas emission above $\sim 1$~keV plays practially no role. For this reason, in treating this problem it is important to use HMXB statistics (at the present epoch) in the soft X-ray energy band, as has been done in this study.

Our conclusion that the Universe could not be significantly heated by X-rays from HMXBs before $z\sim 10$ is similar to that reached by \cite{madfra16}. This agreement is not unexpected. First of all, we based our treatment on the same assumptions about the history of star formation and increased (due to low metallicity) specific X-ray luminosity of the young stellar population in the early Universe. The novelty of our study is that an actually measured (albeit only at $z=0$) integrated luminosity of HMXBs in the soft X-ray band has been used for the first time in the context of the considered problem. \cite{madfra16} used a population synthesis model for these purposes. However, their estimate of the cumulative emissivity of HMXBs at energies below 2~keV is close to ours, which predetermines an agreement of the outcomes of the two studies. An obvious drawback of our study is that we have considered a simple scenario in which the entire Universe is homogeneously filled with cold gas, whereas \citep{madfra16} carried out a more complicated modeling in the context of cosmic reionization, namely took into account that the IGM consists of HII regions and cold gas between them, so that X-ray heating affects only the cold gas phase. However, the differences between the two calculations prove to be barely noticeable (compare the dependencies of the gas temperature and ionization degree on redshift shown in fig.~8 in \citealt{madfra16} with the corresponding curves for our ``optimistic'' scenario in Fig.~\ref{fig:ion_heat_em} above), which is probably mostly related to the fact that the HII volume fraction becomes substantial (more than 10\%) only at $z\la 10$. 

\begin{acknowledgements}
This research has been supported by grant 14-12-01315 from the Russian Science Foundation. 
\end{acknowledgements}

\end{document}